\documentclass[11pt]{article}
\usepackage{feynmf}
\renewcommand{\thefootnote}{\fnsymbol{footnote}}
\newcounter{line}
\def\hph#1{hep-ph{}/{}#1}

\def\etal{\hbox{\it et al.}{}}

\def\ie{\hbox{\it i.e.}{}}
\def\eg{\hbox{\it e.g.}{}}
\def\etc{\hbox{\it etc.}{}}
\def\nn{\hspace{2mm}}
\def\sss{\scriptscriptstyle}
\def\MeV{\mbox{\rm MeV}}
\def\GeV{\mbox{\rm GeV}}
\def\eV{\mbox{\rm eV}}
\def\sleq{\raisebox{-.6ex}{${\textstyle\stackrel{<}{\sim}}$}}

\def\AGUT{{}\;\;\raisebox{.9ex}{$\times$}\raisebox{-.5ex}%
{$\!\!\!\!\!\!\!\!\sss i=1,2,3$} \,(SMG_i \times U(1)_{\sss B-L,i})}%
\def\Tr{\rm{Tr}}

\def\sVEV#1{\left\langle #1\right\rangle}

\def\on#1#2{{\buildrel{\mkern2.5mu#1\mkern-2.5mu}\over{#2}}}
\def\under#1#2{\mathop{\null#1}\limits_{#2}}          

\begin{document}
\begin{titlepage}
\hfill
\vbox{
    \halign{#\hfil        \cr
           NBI-HE-00-42   \cr
           hep-ph/0011062 \cr
           } 
      }  
\vspace*{1mm}
\begin{center}
{\Large {\bf Neutrino mass matrix in Anti-GUT with see-saw mechanism}\\}
\vspace*{10mm}
{\ H.B. Nielsen}\footnote[3]{E-mail: hbech@nbi.dk}
and {\ Y. Takanishi}\footnote[4]{E-mail: yasutaka@nbi.dk}

\vspace*{.5cm} 
{\it The Niels Bohr Institute,\\
Blegdamsvej 17, DK-2100 Copenhagen {\O}, Denmark}\\
%
\end{center}

\begin{abstract}

We have investigated the predictions of neutrino oscillations 
within extended Anti-GUT, based on a large gauge group 
- the Standard Model and an $B-L$ abelian group assumed 
with separate gauge fields coupling to 
each family of quarks and leptons - $\AGUT$.
We take into account corrections concerning a crude way of accounting 
for the number of ways the vacuum expectation values of the Higgs fields 
of the model can be ordered 
in the Feynman diagrams yielding the mass matrix elements. Performing 
these corrections in a balanced way between the charged fermion sector 
and the neutrino sector (case $\rm I$) leads to the previous version of 
our model that was a fit to the MSW small angle solution region. The fit 
considered here is marginally worse that the previous fit inasmuch as it 
predicts a somewhat lower solar neutrino mass square difference 
(which might though be the only way of avoiding the day-night 
exclusion region). The other two 
versions (case $\rm II$ and case $\rm III$) are
characterised by the heaviest of the see-saw neutrino
matrix elements that are composed either of $\nu_{\sss R\mu}$ and 
$\nu_{\sss R\tau}$ or two $\nu_{\sss R\tau}$. These cases $\rm II$ and
$\rm III$ fit the small mixing angle MSW scenario very well.\\

\vskip 2mm \noindent\
\vskip -3mm \noindent\
\end{abstract}
\vskip 10mm

November 2000
\end{titlepage}
\renewcommand{\thefootnote}{\arabic{footnote}}
\setcounter{footnote}{0}
\setcounter{page}{2}
\newpage
\section{Introduction}
\indent

 The latest results of Super-Kamiokande collaboration~\cite{totsuka} 
suggested that the day-night effect spectra disfavour the MSW~\cite{MSW} 
small mixing angle solution (MSW-SMA) at the level of $95\%$ C.L. 
However, it might be that the MSW-SMA could not be excluded by 
the experiments because the measurement of the day-night effect 
is rather difficult. Therefore, we prefer to allow the possibility 
of the MSW-SMA, since we have recently 
predicted the MSW-SMA within extended Anti-GUT~\cite{nt} framework.

It is also important to study whether the Anti-GUT model 
could, after all, predict the MSW large mixing 
solution (MSW-LMA): in our calculations we have
used a technical correction~\cite{douglas}
(``factorial factor correction'') which takes into account 
the number of Feynman diagrams contributing to a
given mass matrix element in a crude statistical way. In fact we used
the parameters from a fit to quarks and charged leptons mass matrices 
without ``factorial factors correction'', while we partly used them 
in the neutrino oscillation calculations. The major aim of the present 
article is to treat the charged fermions and the neutrinos on 
an equal footing with respect to this technical correction. We have
assumed that one of the parameters is fixed to be unity as in earlier 
works~\cite{oldagut}. In this article we will take into account that 
this parameter is not {\it a priori} unity. However, even these smaller 
variations do not bring our model to fit MSW-LMA but still only 
the MSW-SMA domain.

Although our extended Anti-GUT model is rather restricted with respect 
to the choice of charge combinations for the various fermions and even 
for the Higgs fields introduced to break the gauge group assumed 
in order to keep the already achieved good fits, there are a few 
possibilities for playing around. One of the most important 
possibilities is that we can vary the quantum numbers for the 
Higgs field, called $\phi_{\sss B-L}$, the vacuum expectation 
value (VEV) of which is giving the scale of the see-saw particles 
(the right-handed neutrinos) by breaking the 
$(B-L)$ charge (which is assumed to be gauged even in the form 
of one $(B-L)_i$ for each family where $i$ denotes a generation) 
and thereby the overall scale of the neutrino oscillations.  

In the following section, we describe briefly the 
extended Anti-GUT model for the calculation of neutrino 
mass squared differences and the mixing angles. 
Then, in the next section, we put forward the mass matrices.
In section $4$ we describe the ``factorial correction''. Then in 
section $5$ we will present the experimental data and the ``old'' fits 
on the charged lepton and the quark, while we in section $6$ 
give the results of our calculations. Section $7$ contains our 
conclusion and resum{\'e}.

\section{The extended Anti-GUT model}
\begin{figure}[h!]
\vspace{-8mm}
  \begin{center}
\unitlength=1mm
\begin{minipage}{9cm}
\begin{fmffile}{smafig1}
\begin{fmfgraph*}(50,30)
\fmfpen{thick}
\fmfleft{i}
\fmfright{o}
\fmf{plain}{i,i1,i2,i3,i4,i5,i6,i7,i8,i9,i10,i11,i12,i13,i14,i15,%
i16,i17,i18,i19,o,o1}
\fmfdot{i2,i12,i18,i19}
\fmflabel{\raisebox{9mm}{$\phi_{\sss WS}$}}{i5}
\fmflabel{\raisebox{-11mm}{$\hspace{3mm}\sim10^{2}~\GeV$}}{i8}
\fmflabel{\raisebox{9mm}{$\!\!\!\!\!\!\!\phi_{\sss B-L}$}}{i12}
\fmflabel{\raisebox{-11mm}{$\hspace{-3mm}\approx 10^{12}~\GeV$}}{i11}
\fmflabel{\raisebox{8.5mm}{$\hspace{-0.6mm}\downarrow$}}{i17}
\fmflabel{\raisebox{21mm}{$\hspace{-9.3mm}\underbrace{\xi,T,W,\chi}$}}{i18}
\fmflabel{\raisebox{21mm}{$\hspace{4.6mm}\approx 10^{18}~\GeV$}}{i19}
\fmflabel{\raisebox{-12mm}{$\hspace{-.9mm}S\sim 10^{19}~\GeV$}}{o}
\fmflabel{\raisebox{-9mm}{$\hspace{-4.6mm}\nwarrow$}}{o1}
\end{fmfgraph*}
\end{fmffile}
\end{minipage}
\begin{minipage}[b]{45mm}
\vspace{2cm}\vspace*{\fill} 
Fig.~1 \\
\vspace*{\fill} 
\noindent
The energy scale of the vacuum expectation 
values of the Higgs fields within the extended Anti-GUT model.
\end{minipage}
  \label{tab:energie}
  \end{center}
\vspace{-5mm}
\end{figure}
The Anti-GUT model is characterised by
a gauge group which for each family has a set 
of (family specific) gauge fields. In addition there is an 
extra $U(1)$ gauge group called $U(1)_{f}$. The details 
of the couplings of
the latter is largely specified by the requirements 
of the model being free of gauge and mixed anomalies. So each 
generation gets its own system of gauge particles, $\ie$ gauge 
fields come in generations just like the fermions. The 
extension consists in providing each family also with a gauge 
field coupling to the $(B-L)$-charge of that family alone. 
That is to say we postulate for each family\footnote{Really, 
we should say ``proto-family'' because $(1)$ these proto-family 
get mixed before they are identified with the true families $(2)$ 
even the right-handed charm- and top-quarks get permuted.} the 
subgroup of the well-known grand unification group, $SO(10)$, 
which consists only of those generators
that do not mix the different irreducible representation of 
the Standard Model, 
$SMG_{\sss i}\times U(1)_{\sss B-L,i}=SU(3)\times SU(2)\times U(1)\times U(1)$,
where $SMG$ denotes the Standard Model gauge group. The $U(1)_{f}$ that 
seems a strange extra ingredient in the 
``old'' Anti-GUT model - although called for in order to be able 
to fit the quark masses~\cite{low} - is in the extended version 
just a linear combination of $(B-L)_i$
and $y_i/2$ which happen to have separate anomaly cancellations 
for the right-handed neutrinos and for the rest 
(true Standard Model). Since the ``old'' $U(1)_{f}$ quantum charge is a 
linear combination of $(B-L)_2$, $(B-L)_3$, 
$y_2/2$ and $y_3/2$ any anomaly constraint for $U(1)_{f}$ and the
extended model charges must be automatically satisfied. Thus there 
formally can be no hindrance 
for having $U(1)_{f}$ as well as \underline{separate} $U(1)_{\sss B-L,2}$ 
and $U(1)_{\sss B-L,3}$ gauge groups. 
However, such a possibility would imply that there would be a linear 
combination of the charges that would 
decouple from all the fermions and thus should not really 
be considered. In fact, it is a part of the arguments for the 
Anti-GUT model that one decides to ignore gauge fields which 
decouple from all observed 
fermions on the ground that they would have very
little phenomenological relevance.

The Higgs fields in our model are assigned quantum numbers 
under the gauge group ``extended Anti-GUT'' $=\AGUT$ 
which are determined by seeking a fit to the quark and 
lepton spectra including its mixing angles. It should, however, be kept in 
mind that the possibilities for fitting these charge quantum numbers are 
few, since they are only allowed to take quantised values
- analogous to that these quantum numbers for  the quarks and leptons 
only take simple quantised values - and thus the model remains very 
predictive. The model ends up having only $7$ Higgs fields
falling into four classes according to the order of magnitude of the 
expectation values\footnote{The quantum numbers of the seven boson fields 
are shown in Table $1$.}: %
\begin{list}{\it\arabic{line})}{\usecounter{line}}
\item The smallest VEV Higgs field in this model plays the r{\^o}le 
of the Standard Model Weinberg-Salam Higgs field, $\phi_{\sss WS}$, 
with the VEV at the weak scale being $246~\GeV/\sqrt{2}$.
\item The next smallest VEV Higgs field is also alone 
in its class and breaks the common $B-L$ gauge group 
$U(1)$, common to the all the families. This symmetry is supposed 
to be broken (Higgsed) at the see-saw scale as needed for 
fitting the over all neutrino oscillation 
scale. This VEV is of the order of $10^{12}~\GeV$ and called 
$\phi_{\sss B-L}$. This field was not assumed to exist in the 
``old'' Anti-GUT model\footnote{The influence, on the charged 
lepton mass matrix, of the new Higgs field, $\chi$, is only on 
the off-diagonal elements which are remain suppressed and 
do not dominate any masses or mixing angles.}.
\item The next $4$ Higgs fields are called $\xi$, $T$, $W$, 
and $\chi$ and have VEVs of the order of a 
factor $10$ to $50$ under the Planck unit. That means that if intermediate 
propagators have scales given by the Planck scale, as we assume, they 
will give rise to suppression factors of the order $1/10$ each
time they are needed to cause a transition. The field 
$\chi$ was introduced for the purpose of the study of neutrinos 
and was not present in the ``old'' model. 
\item The last one, with VEV of the order of the 
Planck scale, is the Higgs field $S$, which gives little 
or no suppression when it is applied. Therefore a 
transition amplitude in first approximation is
not noticeably suppressed by this field $S$. Thus it
gives rise to ambiguities in the model and its 
presence are not easily distinguished in phenomenology. Only if we take the 
VEV discernibly different from zero and/or make use of the 
``factorial corrections'', there is a possibility to observe 
phenomenological consequences of the field $S$.   
\end{list}

Therefore it is part of the model also that all physics beyond Standard 
Model is in unextended Anti-GUT first a couple of orders of magnitude 
under the Planck scale and in the extended one begins at the see-saw 
scale. So there is pure Standard Model to see-saw scale 
and consequences of the Standard Model such as the GIM 
mechanism~\cite{GIM} of no flavour changing neutral current are 
valid with the correspondingly very high accuracy, by far 
unaccessible to present 
days experiments. 

Since we have non-zero VEVs of scalar fields which are doublets under two 
different ``proto''-generation specific $SU(2)$'s such as 
$S$, $W$, $T$, $\xi$ (while $\phi_{\sss WS}$ has to be doublet 
for an odd number of generations), the mass eigenstates of the quarks and 
leptons do \underline{not} belong to definite 
representations of the ``proto''-generation specific $SU(2)$'s.
There is therefore no hindrance in having non-zero mixing 
angles, but they are of course suppressed corresponding 
to the need for the breaking VEV's being involved.

\begin{table}[!ht]
\caption{All $U(1)$ quantum charges in extended Anti-GUT model. We presented
here the three different possibilities of the $\phi_{\sss B-L}$ 
quantum charges. The symbols for the fermions shall be considered to mean
``proto''-particles. Non-abelian representations are given by a rule 
from the abelian ones, see after Eq.~($2$).}
\vspace{3mm}
\label{Table1}
\begin{center}
\begin{tabular}{|c||c|c|c|c|c|c|} \hline
& $SMG_1$& $SMG_2$ & $SMG_3$ & $U_{\sss B-L,1}$ & $U_{\sss B-L,2}$ & $U_{\sss B-L,3}$ \\ \hline\hline
$u_L,d_L$ &  $\frac{1}{6}$ & $0$ & $0$ & $\frac{1}{3}$ & $0$ & $0$ \\
$u_R$ &  $\frac{2}{3}$ & $0$ & $0$ & $\frac{1}{3}$ & $0$ & $0$ \\
$d_R$ & $-\frac{1}{3}$ & $0$ & $0$ & $\frac{1}{3}$ & $0$ & $0$ \\
$e_L, \nu_{e_{\sss L}}$ & $-\frac{1}{2}$ & $0$ & $0$ & $-1$ & $0$ & $0$ \\
$e_R$ & $-1$ & $0$ & $0$ & $-1$ & $0$ & $0$ \\
$\nu_{e_{\sss R}}$ &  $0$ & $0$ & $0$ & $-1$ & $0$ & $0$ \\ \hline
$c_L,s_L$ & $0$ & $\frac{1}{6}$ & $0$ & $0$ & $\frac{1}{3}$ & $0$ \\
$c_R$ &  $0$ & $\frac{2}{3}$ & $0$ & $0$ & $\frac{1}{3}$ & $0$ \\
$s_R$ & $0$ & $-\frac{1}{3}$ & $0$ & $0$ & $\frac{1}{3}$ & $0$\\
$\mu_L, \nu_{\mu_{\sss L}}$ & $0$ & $-\frac{1}{2}$ & $0$ & $0$ & $-1$ & $0$\\
$\mu_R$ & $0$ & $-1$ & $0$ & $0$  & $-1$ & $0$ \\
$\nu_{\mu_{\sss R}}$ &  $0$ & $0$ & $0$ & $0$ & $-1$ & $0$ \\ \hline
$t_L,b_L$ & $0$ & $0$ & $\frac{1}{6}$ & $0$ & $0$ & $\frac{1}{3}$ \\
$t_R$ &  $0$ & $0$ & $\frac{2}{3}$ & $0$ & $0$ & $\frac{1}{3}$ \\
$b_R$ & $0$ & $0$ & $-\frac{1}{3}$ & $0$ & $0$ & $\frac{1}{3}$\\
$\tau_L, \nu_{\tau_{\sss L}}$ & $0$ & $0$ & $-\frac{1}{2}$ & $0$ & $0$ & $-1$\\
$\tau_R$ & $0$ & $0$ & $-1$ & $0$ & $0$ & $-1$\\
$\nu_{\tau_{\sss R}}$ &  $0$ & $0$ & $0$ & $0$ & $0$ & $-1$ \\ \hline \hline
$\phi_{\sss WS}$ & $0$ & $\frac{2}{3}$ & $-\frac{1}{6}$ & $0$ & $\frac{1}{3}$ & $-\frac{1}{3}$ \\
$S$ & $\frac{1}{6}$ & $-\frac{1}{6}$ & $0$ & $-\frac{2}{3}$ & $\frac{2}{3}$ & $0$\\
$W$ & $0$ & $-\frac{1}{2}$ & $\frac{1}{2}$ & $0$ & $-\frac{1}{3}$ & $\frac{1}{3}$ \\
$\xi$ & $\frac{1}{6}$ & $-\frac{1}{6}$ & $0$ & $\frac{1}{3}$ & $-\frac{1}{3}$ & $0$\\
$T$ & $0$ & $-\frac{1}{6}$ & $\frac{1}{6}$ & $0$ & $0$ & $0$\\
$\chi$ & $0$ & $0$ & $0$ & $0$ & $-1$ & $1$ \\
$^{1]}\;\phi_{\sss B-L}$ & $0$ & $0$ & $0$ & $1$ & $0$ & $1$ \\ 
$^{2]}\;\phi_{\sss B-L}$ & $0$ & $0$ & $0$ & $0$ & $1$ & $1$ \\ 
$^{3]}\;\phi_{\sss B-L}$ & $0$ & $0$ & $0$ & $0$ & $0$ & $2$ \\ 
\hline
\end{tabular}
\end{center}
\end{table}

In our previous article \cite{nt} we formulated the quantum numbers 
of our model with the use 
of the quantum number for the subgroup of the full gauge group called 
$U(1)_{f}$, but it is more elegant without 
this abelian gauge group. The $U(1)_f$ quantum number, $Q_{f}$, used in 
earlier articles - both in extended Anti-GUT (where it can be avoided), 
and in ``old'' Anti-GUT (where there are no right-handed neutrinos 
and thus no anomaly free $(B-L)$'s, so that $U(1)_{f}$ is unavoidable) 
- is related to the quantum numbers of Table $1$ by
\begin{equation}
Q_f = (B-L)_3 - (B-L)_2 + 2\,(\frac{y_2}{2} - \frac{y_3}{2})\nn.
\end{equation}%
\indent
The calculation of the mass matrices in our model consists 
in evaluating 
for each mass matrix element which of our seven Higgs fields are needed to 
provide the difference in quantum numbers 
between the right-handed Weyl components and the fermion in question 
to the left-handed ones. Then, one imagines that the propagator fermions 
all have Planck or fundamental mass scale masses in a Feynman 
diagram which is often a long chain of interactions with the successive 
Higgs fields that were needed. The order of magnitude of the 
diagram is then of the order of the Planck scale multiplied by a 
``suppression factor'' for each Higgs field used. Now we assume that 
there are of order unity random couplings all along the chain and we 
therefor take for the value of the matrix element the product of the 
suppression factors times {\em a random factor of order unity}. These 
factors are taken as random numbers and at the end a 
logarithmic averaging of the resulting mass eigenvalues and mixing 
angles is taken. 

\section{Mass matrices and Higgs quantum numbers %
in the extended Anti-GUT model}
\indent

To write down the mass matrices we need the quantum numbers of the Higgs
fields. However, there is the freedom that we can modify the charge
assignments without too much change in the predictions if we change the
fields by adding to their quantum number assignment the quantum numbers of
$S$. Thus we shall take the point of view that we really do not 
know from the already developed phenomenology of the charged sector 
(quarks and charged leptons) the abelian quantum numbers 
except modulo those of $S$.

To take into account the ambiguity of the choice of the
quantum charges of Higgs fields, due to the VEV of $S$ being 
of order of unity in Planck units, we could parametrise the quantum number 
combinations by integer parameters, $\alpha$, $\beta$, $\gamma$, 
$\delta$, $\eta$ and $\epsilon$ telling the number of additional 
$S$-quantum number combinations counted from some starting value
of the quantum number combinations of the Higgs fields which are 
given in Table $1$.\footnote{The starting values were though not 
the ones of Table $1$ as seen from equation (\ref{eq:zurueckbz}) below.} This 
consideration leads to the following Higgs field quantum 
charges: %
\begin{eqnarray}  
\label{eq:neuql}
S \!&=&\! (\frac{1}{6},-\frac{1}{6},0,-\frac{2}{3},\frac{2}{3},0)\nonumber\\
W \!&=&\! (-\frac{1}{6},-\frac{1}{3},\frac{1}{2},\frac{2}{3},-1,\frac{1}{3}) + \alpha\, S \nonumber\\
T \!&=&\! (-\frac{1}{6},0,\frac{1}{6},\frac{2}{3},-\frac{2}{3},0) + \beta\, S\nonumber\\
\xi \!&=&\! (\frac{1}{6},-\frac{1}{6},0,\frac{1}{3},-\frac{1}{3},0) + \gamma\, S \nonumber\\
\phi_{\sss WS} \!&=&\! (\frac{1}{6}, \frac{1}{2}, -\frac{1}{6},-\frac{2}{3},1,-\frac{1}{3}) + \delta\, S \\
\chi \!&=&\! (0,0,0,0,-1,1) +\eta\, S \nonumber\\
{\rm case~I}{}: \phi_{\sss B-L}\!&=&\!(0,0,0,1,0,1) +\epsilon\,S \nonumber\\
{\rm case~II}{}: \phi_{\sss B-L}\!&=&\!(0,0,0,0,1,1) +\epsilon\,S \nonumber\\
{\rm case~III}{}: \phi_{\sss B-L}\!&=&\!(0,0,0,0,0,2) +\epsilon\,S \nonumber\nn.
\end{eqnarray}

Table $1$ and Eq.~(\ref{eq:neuql}) is simplified by only containing 
the abelian quantum numbers, but we do in fact imagine that in our 
model we have the proto-generation specified by the following rule:
For each ``proto''-generation $i$ the non-abelian 
representations -- of $SU(2)_i$ and $SU(3)_i$  -- are the smallest 
dimension ones obeying the restriction
\begin{equation}
  \label{eq:SMrule}
  d_i/2+t_i/3+y_i/2=0\;({\rm mod}\;1)\nn.
\end{equation}
Here $t_i$ is the triality being $1$ for $\underline{3}$, 
$-1$ for $\overline{\underline{3}}$ and $0$ for $\underline{1}$ or 
$\underline{8}$, respectively. The duality $d_i$ is $0$ for integer 
spin $SU(2)_i$ representations while $1$ for half-integer $SU(2)_i$ spin.

Concerning the see-saw scale determining field $\phi_{\sss B-L}$, 
we have listed three ``promising'' choices of the quantum numbers in addition
to the shifting $S$-quantum numbers choice. With these 
quantum number combination we get the parameterised 
mass matrices as follows, where we simply write $S$, $W$, \etc\ 
instead of $\sVEV{S}$, $\sVEV{W}$, \etc\ for the VEVs: %

\noindent
the uct-quarks:
\begin{equation}
M_U \simeq \frac{\sVEV{\phi_{\sss WS}}}{\sqrt{2}}\hspace{-0.1cm}
\left(\!\begin{array}{ccc}
        S^{\sss1+\alpha-2\beta+2\gamma+\delta}W^{\dagger}T^2(\xi^{\dagger})^2
        & S^{\sss2+\alpha-2\beta-\gamma+\delta}W^{\dagger}T^2\xi 
        & S^{\sss2\alpha-\beta-\gamma+\delta}(W^{\dagger})^2T\xi \\
        S^{\sss1+\alpha-2\beta+3\gamma+\delta}W^{\dagger}T^2(\xi^{\dagger})^3
        & S^{\sss2+\alpha-2\beta+\delta}W^{\dagger}T^2 
        & S^{\sss2\alpha-\beta+\delta}(W^{\dagger})^2T \\
        S^{\sss3\gamma+\delta}(\xi^{\dagger})^3 
        & S^{\sss1+\delta} 
        & S^{\sss-1+\alpha+\beta+\delta}W^{\dagger}T^{\dagger}
                        \nonumber\end{array} \!\right)\label{M_U}
\nonumber\end{equation}\noindent %
the dsb-quarks:
\begin{equation}
\!M_D \simeq \frac{\sVEV{\phi_{\sss WS}}}{\sqrt{2}}\hspace{-0.1cm}
\left (\!\begin{array}{ccc}
        S^{\sss-1-\alpha+2\beta-2\gamma-\delta}W(T^{\dagger})^2\xi^2 
      & S^{\sss-2-\alpha+2\beta-\gamma-\delta}W(T^{\dagger})^2\xi 
      & S^{\sss2-3\beta-\gamma-\delta}T^3\xi \\
        S^{\sss-1-\alpha+2\beta-\gamma-\delta}W(T^{\dagger})^2\xi 
      & S^{\sss-2-\alpha+2\beta-\delta}W(T^{\dagger})^2 
      & S^{\sss2-3\beta-\delta}T^3 \\
        S^{\sss-2-2\alpha+4\beta-\gamma-\delta}W^2(T^{\dagger})^4\xi 
      & S^{\sss-3-2\alpha+4\beta-\delta}W^2(T^{\dagger})^4 
      & S^{\sss1-\alpha-\beta-\delta}WT
                        \end{array} \!\right) \label{M_D}
\nonumber\end{equation}\noindent %
the charged leptons:
\begin{equation}
\!M_E \simeq \frac{\sVEV{\phi_{\sss WS}}}{\sqrt{2}}\hspace{-0.1cm}
\left(\hspace{-0.2 cm}\begin{array}{ccc}
       S^{\sss-1-\alpha+2\beta-2\gamma-\delta}W(T^{\dagger})^2\xi^2 
  & S^{\sss-2-\alpha+2\beta+3\gamma-\delta}W(T^{\dagger})^2(\xi^{\dagger})^3
  & S^{\sss4-\alpha-4\beta+3\gamma-\delta-\eta}WT^4(\xi^{\dagger})^3\chi\\
    S^{\sss-1-\alpha+2\beta-5\gamma-\delta}W(T^{\dagger})^2\xi^5 
  & S^{\sss-2-\alpha+2\beta-\delta}W(T^{\dagger})^2 
  & S^{\sss4-\alpha-4\beta-\delta-\eta}WT^4\chi\\
    S^{\sss2+2\alpha-4\beta-5\gamma-\delta}(W^{\dagger})^2T^4\xi^5 
  & S^{\sss1+2\alpha-4\beta-\delta}(W^{\dagger})^2T^4 
  & S^{\sss1-\alpha-\beta-\delta}WT
                        \end{array} \hspace{-0.2 cm}\right) \label{M_E}
\end{equation}\noindent%
the Dirac neutrinos:
\begin{equation}
\!M^D_\nu \simeq \frac{\sVEV{\phi_{\sss WS}}}{\sqrt{2}}\hspace{-0.1cm}
\left(\hspace{-0.2 cm}\begin{array}{ccc}
        S^{\sss1+\alpha-2\beta+2\gamma+\delta}W^{\dagger}T^2(\xi^{\dagger})^2 
 & S^{\sss2+\alpha-2\beta+3\gamma+\delta}W^{\dagger}T^2(\xi^{\dagger})^3
&S^{\sss2+\alpha-2\beta+3\gamma+\delta-\eta}(W^\dagger)T^2(\xi^\dagger)^3\chi\\
 S^{\sss1+\alpha-2\beta-\gamma+\delta}W^{\dagger}T^2\xi 
& S^{\sss2+\alpha-2\beta+\delta}W^{\dagger}T^2 &
        S^{\sss2+\alpha-2\beta+\delta-\eta}(W^\dagger)T^2\chi\\
 S^{\sss-2+\alpha+\beta-\gamma+\delta+\eta}W^{\dagger}T^\dagger\xi\chi^\dagger
& S^{\sss-1+\alpha+\beta+\delta+\eta}W^{\dagger}T^\dagger\chi^\dagger 
 & S^{\sss-1+\alpha+\beta+\delta}W^{\dagger}T^{\dagger}
                        \end{array}\hspace{-0.2 cm}\right) \label{Md_N}
\nonumber\end{equation}

Now we have to get the right-handed Majorana neutrino mass matrix to be
able to calculate the effective neutrino mass matrix
using the see-saw mechanism~\cite{see-saw1,see-saw2,see-saw3}. However, 
there are
three different choices of the quantum charge of $\phi_{\sss B-L}$, none of 
which give results that can be excluded immediately:
the $(1,3)$-component of the right-handed mass matrix is dominant (called
case $\rm I$), another choice is the $(2,3)$-component is the dominant one 
(called $\rm II$), and third one is the $(3,3)$-component as the 
dominant one (called $\rm III$). In our previous paper, the right-handed
neutrino mass matrix is considered as case $\rm I$.

\noindent\ The right-handed Majorana neutrinos is given in
the case ${\rm I}${} by: 
\begin{equation}
M_R \simeq  \sVEV{\phi_{\sss B-L}}\hspace{-0.1cm}
\left (\hspace{-0.2 cm}\begin{array}{ccc}
S^{\sss-1-\gamma+\eta-\epsilon}\chi^\dagger\xi 
& S^{\sss\eta-\epsilon}\chi^\dagger 
& S^{\sss-\epsilon}  \\
 S^{\sss\eta-\epsilon}\chi^\dagger 
& S^{\sss1+\gamma+\eta-\epsilon}\chi^\dagger\xi^\dagger 
& S^{\sss1+\gamma-\epsilon}\xi^\dagger\\
 S^{\sss-\epsilon}
& S^{\sss1+\gamma-\epsilon}\xi^\dagger 
& S^{\sss1+\gamma-\eta-\epsilon}\chi\xi^\dagger
\end{array} \hspace{-0.2 cm}\right ) \label{Mr_NI}
\end{equation}
\noindent\ In the case ${\rm II}${}:
\begin{equation}
M_R \simeq  \sVEV{\phi_{\sss B-L}}\hspace{-0.1cm}
\left (\hspace{-0.2 cm}\begin{array}{ccc}
S^{\sss-2-2\gamma+\eta-\epsilon}\chi^\dagger\xi^2 
& S^{\sss1+\gamma+\eta-\epsilon}\chi^\dagger\xi 
& S^{\sss1+\gamma-\epsilon}\xi  \\
S^{\sss1+\gamma+\eta-\epsilon}\chi^\dagger\xi 
& S^{\sss\eta-\epsilon}\chi^\dagger 
& S^{\sss-\epsilon}\\
  S^{\sss1+\gamma-\epsilon}\xi 
& S^{\sss-\epsilon} & S^{\sss-\eta-\epsilon}\chi
\end{array} \hspace{-0.2 cm}\right ) \label{Mr_NII}
\end{equation}
\noindent\ In the case ${\rm III}${}:
\begin{equation}
M_R \simeq  \sVEV{\phi_{\sss B-L}}\hspace{-0.1cm}
\left (\hspace{-0.2 cm}\begin{array}{ccc}
S^{\sss-2-2\gamma+2\eta-\epsilon}(\chi^\dagger)^2\xi^2 
& S^{\sss-1-\gamma+2\eta-\epsilon}(\chi^\dagger)^2\xi 
& S^{\sss-1-\gamma+\eta-\epsilon}\chi^\dagger\xi  \\
 S^{\sss-1-\gamma+2\eta-\epsilon}(\chi^\dagger)^2\xi
& S^{\sss2\eta-\epsilon}(\chi^\dagger)^2
& S^{\sss\eta-\epsilon}\chi^\dagger \\
 S^{\sss-1-\gamma+\eta-\epsilon}\chi^\dagger\xi 
& S^{\sss\eta-\epsilon}\chi^\dagger
& S^{\sss-\epsilon}
\end{array} \hspace{-0.2 cm}\right ) \label{Mr_NIII} \nn.
\end{equation}

Note that the quantum numbers of our different Higgs fields 
are not totally independent in so as far as 
there is a relation between the quantum numbers,
\begin{equation}
  \label{eq:neuChiTW}
 \vec{Q}_\chi = 3\,\vec{Q}_W - 9\,\vec{Q}_T + (-6-3\alpha+9\beta+\eta)%
\,\vec{Q}_S \nn,
\end{equation}
and thus the Higgs field combinations needed for a given transition are not
unique, so the choice of the largest contribution for each matrix element
must be selected. To compare with our earlier 
work (except for~\cite{douglas}) and Table $1$, we remark 
that the quantum number combination 
used in that work is obtained by putting
\begin{equation}
  \label{eq:zurueckbz}
\alpha=\beta=1\nn,\nn \gamma=0 \nn,\nn\delta=-1\nn,\nn\eta=0\nn,%
\nn\epsilon=0\nn.
\end{equation}
\begin{figure}[b!]
\vspace{-17mm}
  \begin{center}
\unitlength=1mm
\begin{minipage}{9cm}
\begin{fmffile}{smafig2}
\begin{fmfgraph*}(70,20)
\fmfpen{thin}
\fmfstraight
\fmftopn{t}{9}
\fmfbottomn{b}{9}
\fmf{plain}{b1,b2,b3,b4,b5}
\fmf{phantom}{b5,b6}
\fmf{plain}{b6,b7,b8,b9}
\fmf{phantom}{t1,t2,t3,t4,t5}
\fmf{phantom}{t5,t6}
\fmf{phantom}{t6,t7,t8,b9}
\fmf{plain,tension=3}{t2,b2}
\fmf{plain,tension=3}{t3,b3}
\fmf{plain,tension=3}{t4,b4}
\fmf{plain,tension=3}{t7,b7}
\fmf{plain,tension=3}{t8,b8}
\fmfv{decor.shape=cross,decor.filled=full,decor.size=3thick,%
label=\raisebox{6.5mm}{$W\!\!\!\!\!\!\!$}}{t2}
\fmfv{decor.shape=cross,decor.filled=full,decor.size=3thick,%
label=\raisebox{1.5mm}{$T\!\!\!\!\!\!$}}{t3}
\fmfv{decor.shape=cross,decor.filled=full,decor.size=3thick}{t4}
\fmfv{decor.shape=cross,decor.filled=full,decor.size=3thick}{t7}
\fmfv{decor.shape=cross,decor.filled=full,decor.size=3thick}{t8}
\fmf{dots,tension=4}{t5,t6}
\fmf{dots,tension=1}{b5,b6}
\fmflabel{\raisebox{0mm}{$\!\!\!\!\!\!\!\!\!\!\!\!\!\!\!\!\!\!\phi_{\sss WS}$}}{t5}
\fmflabel{\raisebox{2.1mm}{$\!\!\!\xi$}}{t7}
\fmflabel{\raisebox{5.8mm}{$\!\!\!\!\!\!\!\!\!\!\!\!\!\!\!\!\!\chi$}}{t9}
\end{fmfgraph*}
\end{fmffile}
\end{minipage}
\begin{minipage}[b]{45mm}
\vspace{2cm}\vspace*{\fill} 
Fig.~2 \\
\vspace*{\fill} 
\noindent
One of the typical Feynman diagrams getting mass matrix element. The 
tadpole lines with crosses symbolise VEV of the different Higgs fields.
\end{minipage}
  \label{tab:Froggatt-Nielsen}
  \end{center}
\vspace{-5mm}
\end{figure}
\section{The number of orderings correction for the Feynman diagrams}
\indent

The technical detail called ``factorial correction'' consists in the 
following argument for a Feynman digram counting correction:
the external lines signifying the attachment of the various Higgs 
fields used to make the quantum numbers match, can be put in several 
different sequences along the chain of fermion propagators in the 
diagram (see Fig.~$2$) providing the transition. For all these 
different sequences quite different fermion 
propagators are used. These differently ordered VEV-attachments are 
expected to be statistically independent and should be added with random 
phases. That means that one has a random walk in the complex plane 
(of amplitude values). The average of the numerical square of the 
amplitude for the mass matrix element, say, is getting additive contributions 
from the various diagrams and so goes up linearly with the number of diagrams.
This means then that the amplitude goes as the square root
of the number of diagrams with different orderings 
of the attachments of the different Higgs field symbols 
designating the action of these Higgs field 
VEV's. If we, for example, think of a matrix element for the electron mass,
it turns out that in the model, it is necessary to use the vacuum 
expectation value  for the Weinberg-Salam Higgs field 
$\sVEV{\phi_{\sss WS}}$ once, for the $\xi$ twice, for 
the $T$ twice, for the $W$ once, and for 
the $S$ field $n$ times where $n$ depends on the specific assignment of 
the quantum numbers to the other fields.

The number of orderings in which we can have the attachments of the VEV's
for the electron mass generating diagram (in first approximation)
is then $(6+ n)!/(2!\,2!\,n!)$. This means that we expect 
the amplitude statistically to be 
$\phi_{\sss WS} \xi^2 T^2 W S^n \,\sqrt{(6+n)!/(2!\, 2! \,n!)}$.
When these crude estimates of this correction were taken into account in the
charged quark and lepton fits, it turned out that typically a somewhat 
smaller value of the expectation value for $S$ was called for, say, around 
$1/2$. Also the other VEVs would be somewhat 
changed in the ``improved'' fit including this ``factorial correction'' 
(so called because we have seen that it is square roots of factorials 
that come in).
It is clear that the parameter $\xi$, which roughly plays the r{\^o}le 
of the Cabibbo angle and also explains the ratio of the mass 
of the first- to the second-generation
(as being $\xi^2$), will be smaller with the ``factorial corrections'' 
included in a fit because of the highly suppressed first-generation 
masses, of course, tended to have more fields to permute and thus a larger 
enhancement due the ``factorials'' than, say, the second-generation. Such 
a fit, then, must make $\xi$ smaller with ``factorials'' in order to 
compensate so as to keep the first- to second-generation mass ratio 
fixed. Indeed the $\xi$-VEV without ``factorials'' is fit to 
about $1/10$ whereas with ``factorials'' is instead rather $1/30$.

Here is another little detail: Since the field $S$ has the expectation value 
near unity (in Planck units) having one or several $S$-factors is not 
distinguishable from the fitting. Hence, $S$-factors are not very well
determined from phenomenology. Now the number of $S$-factors 
needed to make a given quantum number shift can be changed 
by modifying the assignment
of the quantum numbers for some of the {\em other} Higgs fields.
Hence one can compensate for the changed quantum number by 
accompanying the other fields 
with a number of $S$-fields, so as to get the the same total transition 
in quantum number. 

\section{Data and fitting of charged sectors}
\indent

Before the results are presented, we should review briefly 
the neutrino experimental 
data~\cite{totsuka,SK,suzukiutakeuchi,chlorine,sage,gallex,gno,BKS,valle}:
the best fit to the Super-Kamiokande atmospheric neutrino data shows
near-maximal mixing angle and 
$\Delta m^2_{\rm atm}\simeq 3.2\times 10^{-3}~\eV^2$, the $90\%$ C.L.
range being $(2-7)\times10^{-3}~\eV^2$. As for the solar neutrino 
data there are four different solutions; the MSW-SMA require 
values of the mass square difference and mixing angle 
that (at $99~\%$ C.L.) lie in the intervals
\begin{eqnarray}
\label{eq:MSW-SMA}
&& 4.0\times 10^{-6}~\eV^2 \; \sleq \; \Delta m_{\odot}^2 \; \sleq \; %
1.0\times 10^{-5}~\eV^2 \nn, 
\nonumber \\  
&& 1.3 \times 10^{-3} \; \sleq \; \sin^22\theta_{\odot} \; \sleq \; %
1.0\times 10^{-2} \nonumber \\
&& (3.25\times 10^{-4} 
 \; \sleq \; \tan^2\theta_{\odot} \; \sleq \; 2.5\times 10^{-3})\nn,\nonumber
\end{eqnarray}%
whereas the MSW-LMA solution is realised in the intervals
\begin{eqnarray}
  \label{eq:MSW-LMA}
&& 7.0\times 10^{-6}~\eV^2 \; \sleq \; \Delta m_{\odot}^2 \; \sleq %
\; 2.0\times 10^{-4}~\eV^2\nn, \nonumber \\
&& 0.50 \; \sleq \; \sin^22\theta_{\odot} \; \sleq \; 1.0 \nn (0.17 %
\; \sleq \; \tan^2\theta_{\odot}  \; \sleq \; 1.0) \nn.\nonumber
\end{eqnarray}%
The LOW solution lies approximately in the region 
\begin{eqnarray}
  \label{eq:LOW}
&& 0.4\times 10^{-7}~\eV^2 \;\sleq\; \Delta m_{\odot}^2 \;\sleq\;  
1.5\times 10^{-7}~\eV^2 \nn, \nonumber \\
&& 0.80 \; \sleq \; \sin^22\theta_{\odot} \; \sleq \; 1.0 %
\nn (0.38 \; \sleq \; \tan^2\theta_{\odot}  \; \sleq \; 1.0)\nn, \nonumber
\end{eqnarray}
and the Vacuum oscillation (VO) in
\begin{eqnarray}
  \label{eq:VO}
&& 5\times10^{-11}~\eV^2 \;\sleq\; \Delta m_{\odot}^2 \;\sleq\;  
5\times10^{-10}~\eV^2 \nn, \nonumber \\
&& 0.67 \sleq\;\sin^22\theta_{\odot} \; \sleq \; 1 %
\nn (0.27 \sleq\; \tan^2\theta_{\odot}  \; \sleq \; 4)\nn. \nonumber
\end{eqnarray}

Since the parameter values resulting from the fits to the charged mass 
matrices for, $\eg$, $\xi$ are of the order of a factor $3$ smaller 
than without the ``factorial correction'', we have 
{\it a priori} an uncertainty of that order if we are not careful to 
treat the neutrino part of the calculation in the same way as the charged 
particle sector. In our previous paper we made partial
``factorial corrections'' for the neutrinos using 
the ``old'' parameters, of which $\xi$ is most important, taken from 
fits to charged masses and mixing angles without ``factorial corrections''. 
One can immediately foresee that making ``factorial corrections'' for both
charged fermions and neutrinos should have the same effect as correcting 
the previous work~\cite{douglas} - very roughly - as if we 
simply decrease the $\xi$-value used for the neutrino predictions
by about a factor $3$.

\begin{table}[!t]
\caption{Fits including ${\cal O}(1)$ factors to quarks and charged leptons . 
The VEV of Higgs fields are measured in the Planck unit. For 
the notation of $\alpha\,,\beta\,, \gamma\,, \delta$, see formulas (\ref{eq:neuql}). }
\begin{displaymath}
\begin{array}{c|rrrr|ccccc||c}
&\alpha & \beta & \gamma & \delta & & \sVEV{W} & \sVEV{T} &
 \sVEV{S} & \sVEV{\xi} & \tilde{\chi}^2 \\ \hline \hline
{\rm i} & -1 & -1 & -1 & 1 & &0.0741 & 0.0635 & 0.487 & 0.0331 & 1.57 \\
{\rm ii}& -1 & 0 & -1 & 1 & &0.0945 & 0.0522 & 0.347 & 0.0331 & 1.41 \\
{\rm iii}& -1 & 0 & 1 & -1 & &0.0857 & 0.0522 & 0.686 & 0.0365 & 1.59 \\
{\rm iv}& -1 & 1 & -1 & 1 & &0.0894 & 0.0525 & 0.756 & 0.0247 & 1.26 \\
{\rm v} & -1 & 1 & 1 & 1 & &0.0945 & 0.0474 & 0.653 & 0.0365 & 1.46 \\
{\rm vi} & 0 & 0 & -1 & 0 & &0.0741 & 0.0810 & 0.286 & 0.0300 & 1.27 \\
{\rm vii}& 0 & 0 & -1 & 1 & &0.0857 & 0.0548 & 0.442 & 0.0347 & 1.40 \\
{\rm viii}& 0 & 0 & 0 & 0 &  &0.0816 & 0.0735 & 0.299 & 0.0331 & 1.37 \\
{\rm ix} & 0 & 1 & 1 & 0 &  &0.0945 & 0.0522 & 0.721 & 0.0331 & 1.62 \\
{\rm x}& 1 & -1 & -1 & 1 & & 0.0857 & 0.0522 & 0.622 & 0.0300 & 1.44 \\
{\rm xi}& 1 & -1 & 1 & 1 & & 0.0900 & 0.0497 & 0.622 & 0.0383 & 1.70 \\
{\rm xii}& 1 & 0 & -1 & 0 & & 0.0900 & 0.0522 & 0.537 & 0.0422 & 1.79 \\
{\rm xiii}& 1 & 0 & -1 & 1 & & 0.0816 & 0.0549 & 0.538 & 0.0331 & 1.64 \\
{\rm xiv}& 1 & 1 & -1 & 0 & & 0.1042 & 0.0522 & 0.346 & 0.0444 & 1.85 \\
\hline
\end{array}
\end{displaymath}
\label{ErgebmitRF}
\end{table}

In~\cite{douglas} several fits to the masses and mixing angles 
for the charged lepton and quark were found using different 
quantum number assignments for the Higgs fields\footnote{%
In the paper~\cite{douglas} the gauge group does not
include the see-saw sector; therefore in fitting the 
neutrino masses and mixing angles using right-handed neutrinos, 
we have to add new gauge fields and Higgs fields which 
spontaneously break the $U(1)_{\sss B-L}$ gauge group; 
$\ie$ we have to introduce 
the additional parameters, namely $\epsilon$ and $\eta$.
They can also be taken in the range $|\epsilon|\le1$ and 
$|\eta|\le1$. See section $3$ for details.} $\xi$, $T$, $W$, 
$\phi_{\sss WS}$ with respect to changing 
them by the quantum number combination of $S$, where we also 
vary $\sVEV{S}$ rather than fixing $\sVEV{S}=1$. The best fitting 
quantum number assignments from formula within the range $|\alpha|\le1$, 
$|\beta|\le1$, $|\gamma|\le1$ and $|\delta|\le1$  are reviewed in 
Table~\ref{ErgebmitRF}. The notation $\tilde{\chi}^2$ in the last column of 
Table~\ref{ErgebmitRF} is misleading with respect to its 
normalisation in as for as it is defined as what $\tilde{\chi}^2$ would 
be if the uncertainty in the logarithms were unity:
\begin{equation}
\label{e12}
\tilde{\chi}^2 = \sum \left[\ln \left(\frac{m}{m_{\mbox{\small{exp}}}} %
\right) \right]^2
\end{equation}
where $m$ are the fitted charged lepton and quark masses and 
mixing angles and $m_{\mbox{\small{exp}}}$ are the 
corresponding experimental values. The Yukawa matrices are 
calculated at the fundamental scale which we take to be 
the Planck scale. We use the first order renormalization 
group equations (RGEs) for the Standard Model to 
calculate the matrices at lower scales. Running masses are calculated 
in terms of the Yukawa couplings at $1~\GeV$. A typical result of 
a fit including averaging over the order of unity $(\ie~{\cal O}(1))$
random numbers (see last of the paragraph of section $2$) is 
represented in Table~\ref{MassenmitFF}.\footnote{Really 
Table $3$ which is copied from Ref. \cite{douglas} is said to 
have ${\cal O}(1)$ factors in distinction to another table in   
that article in which only random phases were used.}
\begin{table}[!t]
\caption{Typical fit including averaging over ${\cal O}(1)$ factors 
with $\alpha=-1$, $\beta=1$, $\gamma=1$ and $\delta=1$. All quark masses 
are running masses at $1~\GeV$ except the top quark mass which 
is the pole mass.}
\begin{displaymath}
\begin{array}{c|c|c}
& {\rm Fitted} & {\rm Experimental} \\ \hline \hline
m_u & 3.1 ~\MeV & 4 ~\MeV \\
m_d & 6.6 ~\MeV & 9 ~\MeV \\
m_e & 0.76 ~\MeV & 0.5 ~\MeV \\
m_c & 1.29 ~\GeV & 1.4 ~\GeV \\
m_s & 390 ~\MeV & 200 ~\MeV \\
m_{\mu} & 85 ~\MeV & 105 ~\MeV \\
M_t & 179 ~\GeV & 180 ~\GeV \\
m_b & 7.8 ~\GeV & 6.3 ~\GeV \\
m_{\tau} & 1.29 ~\GeV & 1.78 ~\GeV \\
V_{us} & 0.21 & 0.22 \\
V_{cb} & 0.023 & 0.041 \\
V_{ub} & 0.0050 & 0.0035 \\
J_{\sss CP} & 1.04\times10^{-5} & 2\!-\!3.5\times10^{-5}\\ \hline \hline
\tilde{\chi}^2 & 1.46 & - \\ \hline 
\end{array}
\end{displaymath}
\label{MassenmitFF}
\end{table}
\setcounter{figure}{2}
\section{Results and discussion}
\indent
\begin{table}[!!!hh]
\begin{center}\vspace{-18.5mm}
\caption{The numerical results of the three different cases, ${\rm I, II}$ and ${\rm III}$
combined with various charged sector fits denoted by small Roman numbers. The best fit, 
case ${\rm III-ix}$ with $\eta=0$ and $\epsilon=-1$, is marked with black bullet. See more detail
in the text.}
\vspace{-1.5mm}
\begin{tabular}{|crc||c|c|c|c|c|c|} \hline
& & & $\under{\tan^2\theta_{\odot}}{[\sin^22\theta_{\odot}]}$ & $\on{}{\frac{\Delta m_{\odot}^2}{\Delta m_{\rm atm}^2}}$ & $\under{\tan^2\theta_{\rm atm}}{[\sin^22\theta_{\rm atm}]}$ & $\under{\tan^2\theta_{e3}}{[\sin^22\theta_{e3}]}$ & $\on{}{\sVEV{\chi}}$\\\hline\hline %
& \multicolumn{2}{|c||}{VO} & $\under{0.23-1.0}{[0.6-1.0]}$ &$\on{}{\approx10^{-7}}$ & & & $-$ \\ \cline{2-5} \cline{8-8} 
$\on{}{\on{}{\rm experi-}}$ & \multicolumn{2}{|c||}{$\on{}{\on{}{\rm SMA}}$} & $\under{(0.33-2.5)\times10^{-3}}{[(1.3-10)\times10^{-3}]}$ & $1.5\,{ +1.5\atop -0.7 }\times10^{-3}$ & $0.485-1.0$  & $\sleq 0.026$ & $-$\\ \cline{2-5}  \cline{8-8} mental& \multicolumn{2}{|c||}{$\on{}{\on{}{\rm LMA}}$} & $\under{0.17-1.0}{[0.5-1.0]}$ & $9.4\,{ +14\atop -6 }\times10^{-3}$ & $[0.88-1.0]$ & $ [\sleq0.1] $ & $-$\\ \cline{2-5}  \cline{8-8}
$\on{}{\on{}{\rm data}}$& \multicolumn{2}{|c||}{$\on{}{\on{}{\rm LOW}}$} & $\under{0.38-1.0}{[0.8-1.0]}$ & $3.1\,{ +11\atop -2.3 }\times10^{-5}$   & & & $-$\\ \hline \hline
\multicolumn{3}{|l|}{\quad{\rm I-xiii}\quad$\eta=0,\;\epsilon=1$} & $7.8\times10^{-4}$ & $6.7\times10^{-5}$ & $1.00$ & $8.0\times10^{-4}$ & $0.018$ \\ \hline
\multicolumn{3}{|l|}{\quad{\rm I-xiii}\quad$\eta=-1,\;\epsilon=1$} & $7.8\times10^{-4}$ & $7.8\times10^{-5}$ & $1.00$ & $7.8\times10^{-4}$ & $0.038$ \\ \hline\hline
\multicolumn{3}{|l|}{\quad{\rm II-ii}\quad$\eta=1,\;\epsilon=0$} & $3.8\times10^{-3}$ & $1.0\times10^{-2}$ & $0.93$ & $8.8\times10^{-4}$ & $0.016$ \\ \hline
\multicolumn{3}{|l|}{\quad{\rm II-vi}\quad$\eta=1,\;\epsilon=0$} & $4.7\times10^{-3}$ & $5.8\times10^{-3}$ & $0.96$ & $1.0\times10^{-3}$ & $0.018$ \\ \hline\hline
\multicolumn{3}{|l|}{\quad{\rm II-ix}\quad$\eta=0,\;\epsilon=1$} & $1.1\times10^{-3}$ & $1.0\times10^{-2}$ & $0.96$ & $2.1\times10^{-4}$ & $0.026$ \\ \hline\hline
\multicolumn{3}{|l|}{\quad{\rm II-v}\quad$\eta=-1,\;\epsilon=1$} & $1.8\times10^{-3}$ & $8.9\times10^{-3}$ & $1.00$ & $2.8\times10^{-4}$ & $0.022$ \\ \hline
\multicolumn{3}{|l|}{\quad{\rm II-ix}\quad$\eta=-1,\;\epsilon=1$} & $1.2\times10^{-3}$ & $9.0\times10^{-3}$ & $1.00$ & $1.9\times10^{-4}$ & $0.022$ \\ \hline
\multicolumn{3}{|l|}{\quad{\rm II-xi}\quad$\eta=-1,\;\epsilon=1$} & $2.1\times10^{-3}$ & $8.4\times10^{-3}$ & $0.95$ & $2.6\times10^{-4}$ & $0.023$ \\ \hline\hline
\multicolumn{3}{|l|}{\quad{\rm II-vi}\quad$\eta=1,\;\epsilon=-1$} & $4.3\times10^{-3}$ & $6.6\times10^{-3}$ & $0.92$ & $1.0\times10^{-3}$ & $0.018$ \\ \hline\hline
\multicolumn{3}{|l|}{\quad{\rm III-v}\quad$\eta=0,\;\epsilon=0$} & $1.1\times10^{-3}$ & $8.8\times10^{-3}$ & $0.93$ &  $6.6\times10^{-4}$ & $0.032$ \\ \hline
\multicolumn{3}{|l|}{\quad{\rm III-viii}\quad$\eta=0,\;\epsilon=0$} &$2.9\times10^{-3}$ & $8.7\times10^{-3}$ & $0.99$ & $2.7\times10^{-4}$ & $0.063$ \\ \hline
\multicolumn{3}{|l|}{\quad{\rm III-xi}\quad$\eta=0,\;\epsilon=0$} & $1.4\times10^{-3}$ & $8.0\times10^{-3}$ & $0.99$ & $7.3\times10^{-4}$ & $0.030$  \\ \hline\hline
\multicolumn{3}{|l|}{\quad{\rm III-v}\quad$\eta=0,\;\epsilon=1$} & $1.2\times10^{-3}$ & $6.8\times10^{-3}$ & $1.00$ & $6.4\times10^{-4}$ & $0.031$ \\ \hline
\multicolumn{3}{|l|}{\quad{\rm III-viii}\quad$\eta=0,\;\epsilon=1$} & $3.0\times10^{-3}$ & $7.8\times10^{-3}$ & $0.95$ & $2.3\times10^{-4}$ & $0.065$ \\ \hline
\multicolumn{3}{|l|}{\quad{\rm III-xi}\quad$\eta=0,\;\epsilon=1$} & $1.5\times10^{-3}$ & $6.3\times10^{-3}$ & $0.98$ & $6.4\times10^{-4}$ & $0.030$ \\ \hline\hline
\multicolumn{3}{|l|}{\quad{\rm III-ii}\quad$\eta=1,\;\epsilon=1$} & $3.6\times10^{-3}$ & $1.0\times10^{-2}$ & $0.92$ &$1.7\times10^{-3}$ & $0.018$ \\ \hline
\multicolumn{3}{|l|}{\quad{\rm III-vi}\quad$\eta=1,\;\epsilon=1$} & $4.4\times10^{-3}$ & $8.0\times10^{-3}$ & $0.96$ & $2.3\times10^{-3}$ & $0.019$ \\ \hline\hline
\multicolumn{3}{|l|}{\quad{\rm III-v}\quad$\eta=-1,\;\epsilon=1$} & $1.5\times10^{-3}$ & $6.3\times10^{-3}$ & $0.97$ & $6.2\times10^{-4}$ & $0.028$ \\ \hline
\multicolumn{3}{|l|}{\quad{\rm III-ix}\quad$\eta=-1,\;\epsilon=1$} & $1.0\times10^{-3}$ &$6.2\times10^{-3}$ & $0.95$ & $4.2\times10^{-4}$ & $0.028$ \\ \hline\hline
\multicolumn{3}{|l|}{\quad{\rm III-ii}\quad$\eta=1,\;\epsilon=0$} & $3.6\times10^{-3}$ & $1.0\times10^{-2}$ & $0.92$ & $2.4\times10^{-3}$ & $0.019$ \\ \hline\hline
\multicolumn{3}{|l|}{\quad{\rm III-v}\quad$\eta=-1,\;\epsilon=0$} & $1.5\times10^{-3}$ & $6.8\times10^{-3}$ & $0.94$ & $6.2\times10^{-4}$ & $0.028$ \\ \hline
\multicolumn{3}{|l|}{\quad{\rm III-viii}\quad$\eta=-1,\;\epsilon=0$} & $3.0\times10^{-3}$ & $7.5\times10^{-3}$ & $1.00$ & $2.5\times10^{-4}$ & $0.130$ \\ \hline
\multicolumn{3}{|l|}{\quad{\rm III-ix}\quad$\eta=-1,\;\epsilon=0$} & $1.0\times10^{-3}$ & $6.7\times10^{-3}$ & $0.94$ & $4.2\times10^{-4}$ & $0.028$ \\ \hline
\multicolumn{3}{|l|}{\quad{\rm III-xi}\quad$\eta=-1,\;\epsilon=0$} & $1.8\times10^{-3}$ & $6.4\times10^{-3}$ & $0.99$ & $6.3\times10^{-4}$ & $0.028$ \\ \hline\hline
\multicolumn{3}{|l|}{\quad{\rm III-v}\quad$\eta=0,\;\epsilon=-1$} & $1.2\times10^{-3}$ & $7.5\times10^{-3}$ & $0.98$ & $6.4\times10^{-4}$ & $0.031$ \\ \hline
\multicolumn{3}{|l|}{$\!\bullet\!\!$\quad{\rm III-ix}\quad$\eta=0,\;\epsilon=-1$} & $8.3\times10^{-4}$ & $5.8\times10^{-3}$ & $0.97$ & $4.3\times10^{-4}$ & $0.035$ \\ \hline
\multicolumn{3}{|l|}{\quad{\rm III-xi}\quad$\eta=0,\;\epsilon=-1$} & $1.5\times10^{-3}$ & $7.8\times10^{-3}$ & $0.97$ & $6.4\times10^{-4}$ & $0.030$ \\ \hline\hline
\multicolumn{3}{|l|}{\quad{\rm III-iii}\quad$\eta=1,\;\epsilon=-1$} & $1.5\times10^{-3}$ & $7.2\times10^{-3}$ & $0.97$ & $4.5\times10^{-4}$ & $0.025$ \\ \hline
\multicolumn{3}{|l|}{\quad{\rm III-ix}\quad$\eta=1,\;\epsilon=-1$} & $1.3\times10^{-3}$ & $7.5\times10^{-3}$ & $0.97$ & $2.8\times10^{-4}$ & $0.039$ \\ \hline\hline
\multicolumn{3}{|l|}{\quad{\rm III-v}\quad$\eta=-1,\;\epsilon=-1$} & $1.5\times10^{-3}$ & $8.2\times10^{-3}$ & $0.99$ & $6.5\times10^{-4}$ & $0.027$ \\ \hline
\multicolumn{3}{|l|}{\quad{\rm III-ix}\quad$\eta=-1,\;\epsilon=-1$} & $9.8\times10^{-4}$ & $7.9\times10^{-3}$ & $1.00$ & $4.5\times10^{-4}$ & $0.027$ \\ \hline
\multicolumn{3}{|l|}{\quad{\rm III-xi}\quad$\eta=-1,\;\epsilon=-1$} & $1.8\times10^{-3}$ & $7.9\times10^{-3}$ & $1.00$ & $6.4\times10^{-4}$ & $0.027$ \\ \hline
\end{tabular}
\end{center}
\end{table}

In Table $4$ we have, in addition to presenting the measured
numbers for the ratio of the mass square differences and the mixing 
angles for neutrinos, given a series of predictions for these 
quantities for various quantum number assignments of the Higgs 
fields in our model. The first column in this Table - below 
the experimental summary part - specify using a capital Roman number 
$\rm I$, $\rm II$ or $\rm III$ the quantum number assignment for the $B-L$
charge breaking field $\phi_{\sss B-L}$. The dominant element of the 
right-handed Majorana mass matrix, is for the well-fitting scenarios, 
always one from the third column (the $\nu_{\sss R\tau}$ column). 
The capital Roman number denotes then the row number of this 
dominant element. Which element dominates depends of course 
on the $\phi_{\sss B-L}$ quantum 
numbers (see equations~(\ref{Mr_NI}), (\ref{Mr_NII}) and (\ref{Mr_NIII})). 
Furthermore, there is a small Roman number 
referring to the quantum number combination chosen from 
Table~\ref{ErgebmitRF} for the Higgs fields 
$\xi$, $T$, $W$ and $\phi_{\sss WS}$ which were already used in the charged 
sector fitting. We only use combinations with good fitting of the charged 
sector. We also have the freedom to choose the quantum numbers of the fields 
only of relevance for the neutrinos $\chi$ and $\phi_{\sss B-L}$ 
by shifting the Higgs fields $S$ as parameterised by $\epsilon$ 
and $\eta$. We have marked the best fit - case ${\rm III-ix}$ with
the parameter choice $\eta =0 $ and $\epsilon=-1$ - with the black bullet
in Table $4$.

The VEV of $\phi_{\sss B-L}$ is irrelevant when we only look for 
the ratio ${\Delta m_{\odot}^2}/{\Delta m_{\rm atm}^2}$ but not 
for the absolute values of these mass square differences. Therefore we 
do not have to fit the VEV of $\phi_{\sss B-L}$. But the $\chi$-field VEV
is fit - by hand - for each of the quantum number combinations considered 
and this VEV relative to the fundamental units ($\ie$~the suppression 
factor really) is presented in the last column.

In spite of the fact that our previous article used only the case where 
$\phi_{\sss B-L}$ had the quantum numbers of the combination 
$\nu_{\sss R e}\nu_{\sss R \tau}$ which we here call case $\rm I$,
we do not, upon using the factorial correction fitted expectation value 
of $\xi$, $T$, $W$, and $\phi_{\sss WS}$ (especially $\xi$) get as good 
fits to the neutrino mass matrices as before. In fact, the case $\rm I$ 
calculations, using various well-fitting assignments in the charged sector, 
consistently, yield predictions for the solar to atmospheric mass 
square difference ratio, ${\Delta m_{\odot}^2}/{\Delta m_{\rm atm}^2}$ 
that are too small by about one order of magnitude compared to the 
MSW-SMA phenomenological value. This is due to that 
we - ignoring factorial corrections and order of unity
details - get $\xi^4$ for this ratio in the case $\rm I$ quantum number 
assignments. With the slightly ``unfair'' treatment of our previous 
article (using $\xi$ ${\it etc.}$ values deduced without use of 
factorials in the charged sectors - in spite of 
having ``factorials'' in the neutrino 
part of the calculation) - we managed to get an increase by a
factor $6$ for this ratio, but that is not sufficient with the 
correctly fitted $\xi$-VEV. We have therefore only presented 
a few examples with case $\rm I$ in Table $4$.

Both case $\rm II$ and case $\rm III$ have many well-fitting 
quantum number assignments associated with them as is seen 
from Table $4$. Generally, the tendency for both 
case $\rm II$ and case $\rm III$ is to predict the 
mass square difference ratio 
${\Delta m_{\odot}^2}/{\Delta m_{\rm atm}^2}$ a bit 
too high by half an order of magnitude. Also the solar neutrino angle is 
predicted a bit too high. 

The case $\rm III$ fits a little better than case $\rm II$
because case $\rm III$ predicts the solar mixing angle a little
better, $\ie$~to be a bit smaller\footnote{Equally good fits to the 
mass square difference ratio, 
${\Delta m_{\odot}^2}/{\Delta m_{\rm atm}^2}=5.8\times10^{-3}$ 
is obtained for solutions ${\rm II-vi}$ and ${\rm III-ix}$, 
but ${\rm III-ix}$ has the solar mixing angle closer to the 
global experimental fit \cite{valle}, 
$\tan^2\theta_{\odot}\simeq 5.8\times10^{-4}$.}.
In all cases our predictions of the solar mixing angle is in the 
region of $1.5\times10^{-3}$. It fits very well for the 
MSW-SMA. It is, however, not compatible with the other solution regions
VO, LMA nor LOW. So it is necessary - unless 
our model is drastically changed - that the MSW-SMA is 
upheld. Otherwise our model disagrees significantly.
\begin{figure}[!t!]
\begin{center}
  \input{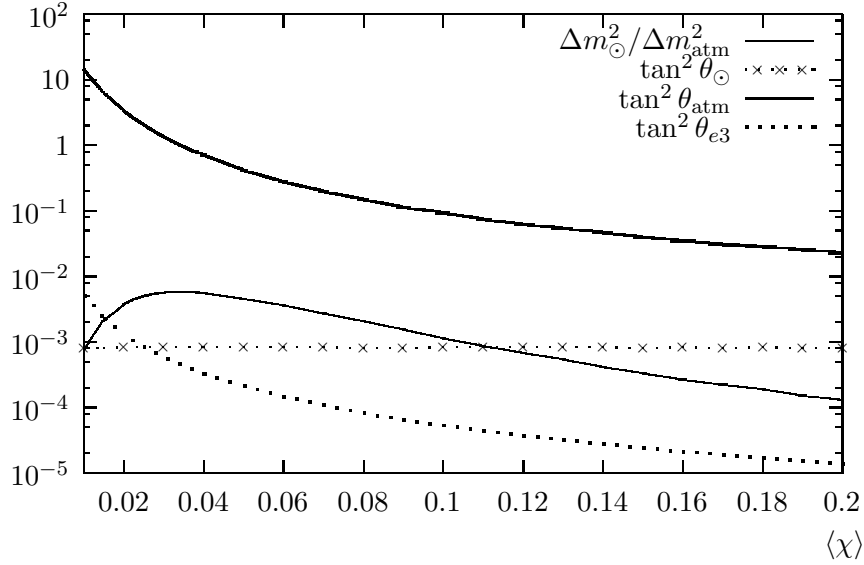}
\label{fig:resutsbest}
\caption{The numerical results in the best 
    case, {\rm III-ix} with $\eta=0$ 
    and $\epsilon=-1$, of the ratio of the 
    solar neutrino mass square difference to that 
    for the atmospheric neutrino oscillation (thin solid line), 
    and the squared tangent of the solar neutrino 
    mixing angle (cross-dotted line), the atmospheric neutrino
    mixing angle (thick solid line) and the mixing 
    angle $\theta_{e3}$ (dotted line) as a function of the 
    VEV of $\chi$ Higgs field. The results are 
    obtained with $100,000$ order of unity random number selections.}
  \end{center}
\end{figure}
Actually, this decrease in $\xi$-value 
insures that the first 
generation matrix elements proportional to $\xi$ or $\xi^2$ have many 
fields and therefore get large ``factorials'' that must be compensated for. 
In Table $2$ we show some numerical results of the most 
suggestive calculations: the combination 
$(\alpha=-1,\,\beta=1,\,\gamma=-1,\,\delta=1)$, $\ie$, case ${\rm iv}$,
giving the very best fit to the quarks and charged leptons supplemented 
with the trivial case $(\ie, \eta=0,\,\epsilon=0)$ gave 
no modification of the already rather simple quantum number 
combinations for $\chi$ and $\phi_{\sss B-L}$ 
in Table $1$. This fitting 
of only $\chi$ to neutrinos (case ${\rm III-iv}$ with $\eta=0$ and 
$\epsilon=0$) gives $\tan^2\theta_{\odot}=7.5\times10^{-3}$ and 
${\Delta m_{\odot}^2}/{\Delta m_{\rm atm}^2}=1.3\times10^{-2}$. 
We also considered some variations in $\eta$ and 
$\epsilon$ in order to study other than the very best fits of the 
charged sector. The main point is that the different choices 
do not yield very different 
fits to the neutrino oscillations and that there are very many good fits
of which a selection - the best mainly - has been put in the table. 

In the examples with case $\rm II$ and also case $\rm III$, we tend 
to get close to the MSW-SMA solution although the solar neutrino 
mass square difference $\Delta m_{\odot}^2$ is predicted a bit to the 
high side of experiments, $\ie$ 
$\Delta m_{\odot}^2/\Delta m_{\rm atm}^2\approx6\times10^{-3}$ 
compared to the experimental 
$\Delta m_{\odot}^2/\Delta m_{\rm atm}^2\approx2\times10^{-3}$. Using 
the best fitting choice of which $S$-quantum numbers are 
added to the Higgs fields 
$(\alpha=0,\,\beta=1,\,\gamma=1,\,\delta=0,\,\eta=0,\,\epsilon=-1)$
gave results just of that character (see also Fig.~$3$):
\begin{eqnarray}
  \label{eq:bestresults}
  \frac{\Delta m_{\odot}^2}{\Delta m_{\rm atm}^2} \!\!&=&\!\! 5.8{+30\atop-5}\times10^{-3}\\
  \tan^2\theta_{\odot}\!\!&=&\!\! 8.3{+21\atop-6}\,\times10^{-4}\nn.
\end{eqnarray}
The deviation from the experimental data of our predictions 
should be considered within uncertainly limits. We interpret 
our ``being uncertain by order unity'' as meaning a relative 
uncertainty $\pm64\%$ for masses and mixing angles. Note 
that for the mass square difference ratio 
$\Delta m_{\odot}^2/\Delta m_{\rm atm}^2$, which involves 
four mass factors, the uncertainly is larger by a factor 
$2\,\sqrt{2}\,\approx\,3$ counted in logarithmic uncertainty meaning 
an uncertainty of the order ${ +511\atop-84} \,\%$. But for the squared 
mixing angles the uncertainty is ``only'' $\pm\,2\times64\,\%$ meaning 
${ +260\atop-72}\,\%$. With these uncertainties the predictions
are in good agreement with experiment.

\subsection{Crude calculation of the mass square difference}
\indent

The results of the see-saw mechanism is that the light neutrino 
masses are quadratic in the Dirac masses and inversely proportional 
to the heavy right-handed Majorana masses:
\begin{equation}
  \label{eq:see-saw}
  M_{\rm eff} \! \approx \! M^D_\nu\,M_R^{-1}\,(M^D_\nu)^T\nn.
\end{equation}

The mass square difference ratio, 
${\Delta m_{\odot}^2}/{\Delta m_{\rm atm}^2}$, from 
equations (\ref{Md_N}), (\ref{Mr_NIII}) and (\ref{eq:see-saw}) 
can be approximately calculated (estimating by rules like
$\lambda_a\,a + \lambda_b\,b\;\simeq \sqrt{a^2+b^2}$, where
$\lambda_a$ and $\lambda_b$ are order of unity random numbers) in the 
best case,~$\ie$,~$\rm III-ix$ with $\eta=0$ and $\epsilon=-1$, as follows:
\begin{eqnarray}
  \label{eq:Meffapprox}
&& M_{{\rm eff}}{}_{\hspace{-0.4mm}\big|}{}_{\textrm{{\tiny\begin{tabular}{l}%
\hspace{-1.8mm} case~{\rm III-ix}\\\hspace{-1.6mm}%
$\eta=0,\,\epsilon=-1$\end{tabular}}}} \!\!\!\!\!\!\simeq %
\frac{\sVEV{\phi_{\sss WS}}^2\,S^2\,W^2 T^2 \chi^2\,{\sVEV{\phi_{\sss B-L}}}^2}{{\rm det} M_{R}}
\overbrace{\left (\hspace{-0.2 cm}\begin{array}{ccc}
6 \sqrt{35}\,S\,T\,\xi^2 & 60 \sqrt{14}\,S^3\,T\,\xi^3 
& 60\sqrt{154}\,S^3\,T\,\xi^3\,\chi\\
6 \sqrt{35}\,S^2\,T\,\xi & 2\sqrt{3}\,T & 2\sqrt{15}\,T\,\chi \\
6\sqrt{70}\,S^2\,\xi\,\chi & 2\sqrt{6} \chi & \sqrt{6}
\end{array} \hspace{-0.2 cm}\right )}^{{\rm part~of}~M_\nu^D} \nonumber\\
&& \;\;\;\times \hspace{0.3cm}
\underbrace{\left(\hspace{-0.2 cm}\begin{array}{ccc}
\sqrt{33}/2 & 2\sqrt{6}\,\xi & 3\sqrt{21}/2\,\chi\xi \\
2\sqrt{6}\,\xi & \sqrt{246}\,\xi^2 & 12\sqrt{5}\,\chi\,\xi^2 \\
3\sqrt{21}/2\,\chi\xi &12\sqrt{5}\,\chi\,\xi^2 & 3\sqrt{165}\,\chi^2\,\xi^2
\end{array} \hspace{-0.2 cm}\right )}_{{\rm part~of}~M^{-1}_R}
\underbrace{\left (\hspace{-0.2 cm}\begin{array}{ccc}
6 \sqrt{35}\,S\,T\,\xi^2 
& 6 \sqrt{35}\,S^2\,T\,\xi
& 6\sqrt{70}\,S^2\,\xi\,\chi \\
 60 \sqrt{14}\,S^3\,T\,\xi^3 
&  2\sqrt{3}\,T 
& 2\sqrt{6} \chi \\
60\sqrt{154}\,S^3\,T\,\xi^3\,\chi
& 2\sqrt{15}\,T\,\chi 
& \sqrt{6} 
\end{array} \hspace{-0.2 cm}\right )}_{{\rm part~of}~(M_\nu^D)^T} \nonumber \\
&& \hspace{1cm}\approx %
\frac{\sVEV{\phi_{\sss WS}}^2\,S^2\,W^2 \,T^2\,\chi^2\,\xi^2\,{\sVEV{\phi_{\sss B-L}}}^2}{{\rm det} M_{R}}
\hspace{-0.1cm}
\left (\hspace{-0.2 cm}\begin{array}{ccc}
\sim\! 0 &\sim\! 0  &\sim\! 0 \\
\sim\! 0 & 1103\,T^2 & 1391\,T\,\chi\\
\sim\! 0 &  1391\,T\,\chi & 2206\,\chi^2
\end{array} \hspace{-0.2 cm}\right )\nn,
\end{eqnarray}
\noindent
where we have used the Higgs fields $S$, $W$, $T$, $\xi$, $\chi$ 
and the fields $S^\dagger$, $W^\dagger$, $T^\dagger$, 
$\xi^\dagger$, $\chi^\dagger$ (with opposite quantum 
charges) equivalently because the non-supersymmetric model 
is considered here. Moreover we have used the numerical value 
$S^2\sim1/2$ and the relation $T\sim\chi$ to obtain the 
large atmospheric mixing angle. It is important that the 
dominant matrix elements are not uncorrelated. The 
determinant of the dominant $2$-by-$2$ 
subgroups of $M_{\rm eff}$ must be calculated by multiplying 
the $2\times2$ determinant of the relevant 2-by-2
sub-matrices of the three matrices, $M_\nu^D$, $M^{-1}_R$ and $(M_\nu^D)^T$:
\begin{eqnarray}
  \label{eq:dettr}
{\rm det}M_{{\rm eff}}{}_{\hspace{-0.4mm}\bigg|}{}_{\textrm{{\tiny\begin{tabular}{l}\hspace{-2.1mm} case~{\rm III-ix}\\\hspace{-2.1mm}$\eta=0,\,\epsilon=-1$\\\hspace{-2.1mm}$(2,3)\times(2,3)$ block\end{tabular}}}} \!\!&=&\!\!
{\rm det}{M^D_\nu}{}_{\hspace{-0.4mm}\big|}{}_{\textrm{{\tiny\begin{tabular}{l}\hspace{-2.1mm} $(2,3)\times(1,2)\,{\rm block}$\\\hspace{-2.1mm}$\eta=0,\,\epsilon=-1$\end{tabular}}}}\;
{\rm det}{M^{-1}_R}{}_{\hspace{-0.4mm}\big|}{}_{\textrm{{\tiny\begin{tabular}{l}\hspace{-2.1mm} $(1,2)\times(1,2)\,{\rm block}$\\\hspace{-2.1mm}$\eta=0,\,\epsilon=-1$\end{tabular}}}} \;\\
 \!\!&& \times\;\;\; {\rm det}{( (M^D_\nu)^T )}{}_{\hspace{-0.4mm}\big|}{}_{\textrm{{\tiny\begin{tabular}{l}\hspace{-2.1mm} $(1,2)\times(2,3)\,{\rm block}$\\\hspace{-2.1mm}$\eta=0,\,\epsilon=-1 $\end{tabular}}}}\nonumber\\
\!\!&\sim&\!\! 3\times10^{6}\;\frac{S^8\,W^4\,T^6\,\chi^6\,\xi^4\sVEV{\phi_{\sss WS}}^4\,\sVEV{\phi_{\sss B-L}}^4}{({\rm det}M_R)^2} \\
({\rm Tr}M_{{\rm eff}})^2{}_{\hspace{-0.4mm}\bigg|}{}_{\textrm{{\tiny\begin{tabular}{l}\hspace{-2.1mm} case~{\rm III-ix}\\\hspace{-2.1mm}$\eta=0,\,\epsilon=-1$\\\hspace{-2.1mm}$(2,3)\times(2,3)$ block\end{tabular}}}}\!\!&\sim&\!\! 6\times10^{6}\;\frac{S^4\,W^4\,T^4\,\chi^8\,\xi^4\sVEV{\phi_{\sss WS}}^4\,\sVEV{\phi_{\sss B-L}}^4}{({\rm det}M_R)^2}
\end{eqnarray}
So we get
\begin{equation}
  \label{eq:meffresulappro}
  \sqrt{\frac{\Delta m_{\odot}^2}{\Delta m_{\rm atm}^2}} \approx \frac{{\rm det}M_{\rm eff}}{({\rm Tr}M_{\rm eff})^2}\,\sim\,\frac{S^4}{2}\,\sim\,\frac{1}{8}
\end{equation}
Actually, the terms coming from the $(1,1)$-component of the $M^{-1}_R$ 
are somewhat dominant due to the factorial corrections, so that the 
dominant $2$-by-$2$ subgroup, the $(2,3)\times(2,3)$ of $M_{\rm eff}$, 
is to first approximation degenerate.
\section{Conclusion and resum{\'e}}
\indent

We have considered a series of extensions so as to include of 
neutrinos of the earlier Anti-GUT fit to the charged quarks 
and lepton masses by introduction of two more Higgs field, $\chi$ 
and $\phi_{\sss B-L}$.

To fit the atmospheric neutrino oscillation experimental data
it is necessary to fix the VEV of the Higgs field, $\sVEV{\chi}$, 
to be in first approximation $\chi\sim T$ in order to arrange 
the atmospheric mixing angle to be of order unity. We should therefore 
not consider the agreement of the atmospheric mixing angle 
$\theta_{\rm atm}$ as one of the significant predictions
of this model: we have, however, managed, at least, by such a fitting 
with $\chi$, to obtain a large mixing angle; $\theta_{\rm atm}$, 
$\ie$, the atmospheric mixing angle is essentially 
an input parameter in this model.

The success of this model should be rather judged from the following 
predictions:
\begin{list}{\it\arabic{line})}{\usecounter{line}}
\item the solar mixing angle which comes out proportional to $\xi$ which
in turn deviates only by ``factorials'' and order of unity 
from the Cabibbo angle.
\item the mass square difference ratio 
${\Delta m_{\odot}^2}/{\Delta m_{\rm atm}^2}$~.
\item the electron- and tau-neutrino mixing angle being small 
enough not to be in conflict with the CHOOZ measurements 
$(U_{e3}=\sin\theta_{e3}\le0.16)$.
\end{list}

The first of these predictions is the major reason that 
our model in the present form ($\ie$ with the present quantum 
number assignments only varied within the limits considered
in this article) is \underline{only} compatible with 
the \underline{MSW-SMA} solar neutrino solution. The smallness 
of the Cabibbo angle results in our solar mixing angle being so 
small that it would stress the model drastically to seek to 
fit one of the series of large solar mixing angle fitting 
regions such as LMA, LOW or VO. 

Within the theoretical uncertainly inherent in our predictions 
being only of order unity - even if taken as specified 
in Ref.~\cite{douglas} with $64\%$ being one standard 
deviation - our model fits the data perfectly provided that
we allow the MSW-SMA solution as the (possibly) valid one in 
spite of the fact that the day-night effect disfavour this solution
by two standard deviation: Extracting from 
the best fit, ${\rm III-ix}$ with $\eta=0\nn, \epsilon=-1$, the result 
may be stated as 
\begin{eqnarray}
  \label{eq:kekka}
  \tan\theta_{\odot}\!&\approx &\!\xi\approx\frac{1}{8}\theta_{\rm c}\nn,\\
  \sqrt{\frac{\Delta m_{\odot}^2}{\Delta m_{\rm atm}^2}}\!&\approx &\!\frac{1}{13} \nn.
\end{eqnarray}
The collective fit of both the neutrinos and the charged sector 
with $6$ fitted VEVs, one of which $\sVEV{S}$ is 
still close to unity and with the $\sVEV{\phi_{\sss WS}}$ determined from 
the weak interaction Fermi constant, is so good order of magnitude-wise 
that it should perhaps be considered perfect.

Since our mass eigenvalues are not tightly degenerate, it is not likely 
- or rather it is impossible - that the renormalization group 
running should be very important as corrections to our predictions.
In fact we expect them to have almost no influence to our 
order of unity accuracy for mass ratios of particles within the
same group such as, say, the (left) neutrinos. 
In fact, in the charged sector, the renormalization group 
mainly just corrects the 
masses of the quarks by a factor $3$ or so compared to the 
corresponding charged leptons.

One could perhaps complain though that we have had too many relatively 
complicated quantum number assignments which, although only discrete 
choices, may be too many possibilities to make our fit convincing. 
The Anti-GUT model with its very many gauge fields can also be 
complained about as being too 
complicated, but here it should be pointed out that indeed it should 
rather gain its reason for being considered by 
being the largest gauge group transforming the Standard Model 
fermions plus the see-saw neutrinos non-trivially among themselves 
without unifying irreducible representations of the Standard Model.

\subsection{Quantum number system of fitting results}
\indent

Presumably the honest motivation for a fit as in the present 
work is that we have a major part of the fitting going on by 
fitting the discrete 
representation choice for the Higgs fields. The best fit has its Higgs 
field (abelian) quantum numbers listed in Table $5$. This Table, in
this sense, represents quantum numbers that are derived/inspired 
from experiment. One can then imagine 
using this inspiration to look for 
regularities suggestive of the model beyond the Standard Model.
\begin{table}[!!t]
\caption{All $U(1)$ quantum charges of the best fit in extended 
Anti-GUT model. The VEV are presented in the Planck unit. Non-abelian 
representations are given by a rule from the abelian ones, see after 
Eq.~($2$).}
\vspace{2mm}
\label{bestqunatum}
\begin{center}
\begin{tabular}{|c||c|c|c|c|c|c|c|} \hline
& $SMG_1$& $SMG_2$ & $SMG_3$ & $U_{\sss B-L,1}$ & $U_{\sss B-L,2}$ & $U_{\sss B-L,3}$ & VEV\\
 \hline\hline
$\phi_{\sss WS}$ & $\frac{1}{6}$ & $\frac{1}{2}$ & $-\frac{1}{6}$ & $-\frac{2}{3}$ & $1$ & $-\frac{1}{3}$ & $\sim10^{-17}$\\
$S$ & $\frac{1}{6}$ & $-\frac{1}{6}$ & $0$ & $-\frac{2}{3}$ & $\frac{2}{3}$ & $0$ & $0.721$\\
$W$ & $-\frac{1}{6}$ & $-\frac{1}{3}$ & $\frac{1}{2}$ & $\frac{2}{3}$ & $-1$ & $\frac{1}{3}$ & $0.0945$\\
$\xi$ & $\frac{1}{3}$ & $-\frac{1}{3}$ & $0$ & $-\frac{1}{3}$ & $\frac{1}{3}$ & $0$ & $0.0331$\\
$T$ & $0$ & $-\frac{1}{6}$ & $\frac{1}{6}$ & $0$ & $0$ & $0$ & $0.0522$\\
$\chi$ & $0$ & $0$ & $0$ & $0$ & $-1$ & $1$ & $0.0345$\\
$\phi_{\sss B-L}$ & $-\frac{1}{6}$ & $\frac{1}{6}$ & $0$ & $\frac{2}{3}$ & $-\frac{2}{3}$ & $2$ & $\sim10^{-8}$\\ \hline
\end{tabular}
\end{center}
\end{table}
It is easy to estimate the VEV for $\phi_{\sss B-L}$ needed in the fit by 
using, say, the example of section $6.1.$ Inserting
\begin{equation}
  {\rm det}M_R \approx 72\; S^3\,\chi^4\,\xi^2\,{\sVEV{\phi_{\sss B-L}}}^3\\
\label{eq:detMR}  
\end{equation}
into equation $(19)$ yields 
\begin{equation}
{\Delta m_{\rm atm}^2}\simeq({\Tr}\,M_{\rm eff})^2 \simeq 1.2\times10^{3}\;
\frac{W^4\,T^4\,\,{\sVEV{\phi_{\sss WS}}}^4}{S^2\,\,{\sVEV{\phi_{\sss B-L}}}^2}\nn.
\end{equation}
Then from equation (\ref{eq:detMR}) together with the atmospheric neutrino
oscillation data we get 
\begin{equation}
  \label{eq:phinumber}
  \sVEV{\phi_{\sss B-L}} \sim 6.7\,\times\,10^{11}~\GeV\nn.
\end{equation}

We present here the order of magnitude of the right-handed
Majorana neutrino masses from the mass matrix, $M_R$, of the 
best fitting case, ${\rm III-ix}$ with $\eta=0$ and 
$\epsilon=-1$ including factorial corrections:
\begin{eqnarray}
  \label{eq:MajoranaMassen}
  M_{R{\sss\nu_{1}}} &\approx& 3\,\times\,10^{6}~\GeV\nn,\nonumber\\
  M_{R{\sss\nu_{2}}} &\approx& 4\,\times\,10^{8}~\GeV\nn,\\
  M_{R{\sss\nu_{3}}} &\approx& 1.4\,\times\,10^{11}~\GeV\nn. \nonumber
\end{eqnarray}

\subsection{Theoretical lesson}
\indent

It is highly remarkable that we obtain the rather good fit of the mass
square difference ratio, 
${\Delta m_{\odot}^2}/{\Delta m_{\rm atm}^2}$, in the two 
well-fitting cases $\rm II$ and $\rm III$:
were it not for the, in principle, order unity factors coming out of
this ratio, it should have simply been unity: 
${\Delta m_{\odot}^2}/{\Delta m_{\rm atm}^2}\approx 1$ since the Higgs 
VEV's drop out except for the Higgs field $S$. The remaining in principle 
order unity ratio comes out of the way 
the various numbers are combined in the calculations of
inverse matrices and matrix products and the ``factorial corrections''
which are due to diagram counting and the field $S$. 

All of the numbers that are not of order unity in our model, namely the Higgs
field VEV (in Planck units) $\xi,\,\chi\,\ldots$, except $S$ which has anyway 
$\sVEV{S}\approx1$, drop out of this ratio. This is the way that our 
model in cases $\rm II$ and $\rm III$ solves one of the 
at first sight almost hopeless problems with data: how,
in fitting the large atmospheric mixing angle, to get 
the hierarchically looking large mass square difference 
ratio between the middle and the heaviest eigenmass neutrinos 
when the highest mass has to mix strongly (of order unity) with 
other eigenstates.

It is, however, indeed possible to interpret this hierarchical mass ratio as 
really due to a hierarchical small VEV as case $\rm I$ does, where this ratio 
goes as $\xi^4$. However, case $\rm I$, does not fit well precisely 
because it makes this ratio too small - too hierarchical.

\subsection{Our qualitative predictions for future neutrino physics}
\indent

We summarise here our predictions for future neutrino experiments
assuming that this model were/is right: %
\begin{list}{\it\arabic{line})}{\usecounter{line}}
\item It should be SMA-MSW that becomes the best fit (the day-night effect 
should be tested by the SNO detector~\cite{petcov}).
\item It is good that the atmospheric mixing angle is of order unity 
but it should \underline{not} turn out to be very close to 
$\tan^2\theta_{\rm atm}\approx 1$, 
$\ie$~$\tan^2\theta_{\rm atm}=1+{\cal O}(\rm lower~order)$. There 
are namely models of a similar spirit (but not our most 
promising fits) which behave this way - for example
when a pair of off-diagonal elements of $M_{\rm eff}$ dominate.
\item KamLAND should not be able to see any neutrino oscillation 
because of SMA-MSW.
\item Since this model only contains three light neutrinos, there is 
no place for the LSND effect~\cite{LSND}.
\end{list}
It is important to confront our model with baryon number production 
in Big Bang which in models like ours with pure Standard Model 
at the weak scale must get the baryon number as a $B-L$ contribution 
from some other scale. This model is of the see-saw type 
with the scale given by the neutrino oscillation scale. After crude 
estimation of the Baryogenesis, we found that this model predicts 
the right range as Ref.~\cite{bayo1,bayo2} studied. Since we 
do not have SUSY in our model - preferably at least - we do not 
have to worry about gravitino problems~\cite{yanagidagroup}. We 
can say that they simply do not exist at all or only at such 
high energy scales 
that they are totally irrelevant. Another potential 
problem~\cite{uppsala,NTbaryo} 
is the wash out of the $B-L$ excess. It is diminished compared 
to the typical lightest see-saw masses considered by \cite{bayo2} because
of our relatively low lightest see-saw mass, $3\,\times\,10^{6}~\GeV$
(see equation (\ref{eq:MajoranaMassen})).

\subsection{The ``efficiency'' of our model}

The number of measured quantities, which are predictable with this model,
quark and charged lepton masses and their mixing angles (three 
Cabibbo-Kobayashi-Maskawa mixing matrix angles) containing Jarlskog 
triangle area, $J_{\sss CP}$ and also two mass square differences and 
the three mixing angles for the neutrino oscillation, Baryogenesis and 
neutrinoless double beta decay~\cite{NTfutur}, is 20.

Our model predicted successfully all these quantities using 
only six parameters\footnote{The VEV of Weinberg-Salam we have 
not counted as a parameter because of its relation to the 
Fermi constant.} which we fit -- VEVs of additional Higgs 
fields: the genuine number
of the predicted parameters is thus 14. However we have taken into 
the predictions two quantities namely, $\tan^2\theta_{e 3}$ and 
``effective'' Majorana neutrino mass for which only experimental 
upper bounds exist. However we must emphasise again that 
strictly speaking we have almost ``infinite'' input parameters
but since we only care for order of magnitude results, we should not count
couplings being order of unity. 

The assumptions of quantum numbers are a bit arbitrary 
or fitted rather and a bit complicated although still 
discrete. But otherwise it is a very reasonable and
expected type of assumptions that are used, $\eg$,~the major 
assumption of all couplings at ``Planck scale'' being of order 
unity is philosophically one of the easiest to get. It is 
(almost) just use of usual dimensionality argumentation. That 
the only deviation from everything being of order unity comes 
from the VEVs may find a bit of support in the well-known 
fact that in super-conductors, VEVs tend often to be very small 
on the atomic physics scale expected {\it a priori}.

\section*{Acknowledgements}
We wish to thank W.~Buchm{\"u}ller, C.D.~Froggatt, S.~Lola, 
M.~Pl{\"u}macher and T.~Yanagida for useful discussions. 
We thank D.~L.~Bennett for reading the manuscript.
We wish to thank the Theory Division of CERN for 
the hospitality extended to us during our visits where most part 
of this work was done. H.B.N. thanks the EU commission for 
grants SCI-0430-C (TSTS), CHRX-CT-94-0621, INTAS-RFBR-95-0567 
and INTAS 93-3316(ext). Y.T. thanks the Scandinavia-Japan 
Sasakawa foundation for grants No.00-22. 


\begin{thebibliography}{99}
%
\bibitem{totsuka}
Y.~Totsuka, talk at 8th International Conference on 
Supersymmetries in Physics (SUSY2K),  26 June - 1 July 2000, 
CERN, Geneva, Switzerland.
\bibitem{MSW} 
L.~Wolfenstein, Phys.\ Rev.\  {\bf D17} (1978) 2369; S.~P.~Mikheyev %
and A.~Yu~Smirnov, Sov.\ J.\ Nucl.\ Phys.\ {\bf 42} (1986) 913.
%
\bibitem{nt}
H.B.~Nielsen and Y.~Takanishi, Nucl.\ Phys.\  {\bf B588} (2000) 281.
%
\bibitem{douglas}
C.D.~Froggatt, H.B.~Nielsen and D.J.~Smith, in progress.
%
\bibitem{oldagut}
see for example: C.~D.~Froggatt, H.~B.~Nielsen and D.~J.~Smith,
Phys.\ Lett.\  {\bf B385} (1996) 150; D.L.~Bennett, C.D.~Froggatt 
and H.B.~Nielsen, Proceeding 
paper of the APCTP-ICTP Joint International Conference (AIJIC 97) on Recent 
Developments in Nonperturbative Quantum Field Theory, Seoul, 
Korea, 26-30 May 1997.
%
\bibitem{low} C.~D.~Froggatt, G.~Lowe and H.~B.~Nielsen, 
Phys.\ Lett.\  {\bf B311} (1993) 163; C.~D.~Froggatt, G.~Lowe 
and H.~B.~Nielsen, Nucl.\ Phys.\  {\bf B414} (1994) 579; C.~D.~Froggatt,
G.~Lowe and H.~B.~Nielsen, Nucl.\ Phys.\  {\bf B420} (1994) 3.
%
\bibitem{GIM}
S.~L.~Glashow, J.~Iliopoulos and L.~Maiani, Phys.\ Rev.\ D {\bf 2} 
(1970) 1285.
%
\bibitem{see-saw1}
M.~Gell-Mann, P.~Ramond
and R.~Slansky in Supergravity, Proceedings of the Workshop at
Stony Brook, NY (1979), eds. P.~van Nieuwenhuizen and D.~Freedman
(North-Holland, Amsterdam, 1979).
\bibitem{see-saw2}
T.~Yanagida, in Proceedings of the Workshop on Unified
Theories and Baryon Number in the Universe, Tsukuba, Japan (1979), eds.
O. Sawada and A. Sugamoto, KEK Report No. 79-18.
\bibitem{see-saw3}
R.~N.~Mohapatra and G.~Senjanovi{\'c}, Phys.\ Rev.\ Lett.\ {\bf 44} (1980) 912.
\bibitem{SK}
Y. Fukuda {\it et al.}, Super-Kamiokande Collaboration, 
Phys.\ Lett.\  {\bf B467} (1999) 185; H. Sobel, 
talk at  XIX International Conference on Neutrino Physics 
and Astrophysics, Sudbury, Canada, June 2000; T. Toshito, 
talk at the XXXth International Conference on High Energy 
Physics, July 27 - August 2, 2000 (ICHEP 2000) Osaka, Japan.
%
\bibitem{suzukiutakeuchi} 
Y. Suzuki, talk  at XIX International Conference on Neutrino Physics 
and Astrophysics, Sudbury, Canada, June 2000; T. Takeuchi, talk 
at the XXXth International Conference on High Energy 
Physics, July 27 - August 2, 2000 (ICHEP 2000) Osaka, Japan.
%
\bibitem{chlorine} B. T. Cleveland {\it et al.}, 
Astrophys. J. {\bf 496} (1998) 505;  R. Davis, Prog. Part. 
Nucl. Phys. {\bf 32} (1994) 13; K. Lande, talk at 
XIX International Conference on Neutrino Physics and 
Astrophysics, Sudbury, Canada, June 2000. 
%
\bibitem{sage} SAGE Collaboration, J. N. Abdurashitov {\it et al.},
Phys. Rev. {\bf C60} (1999) 055801; V. Gavrin, talk at 
XIX International Conference on Neutrino Physics and Astrophysics,
Sudbury, Canada, June 2000.
%
\bibitem{gallex} GALLEX Collaboration, W.~Hampel {\it et al.},
Phys.\ Lett.\ {\bf B447} (1999) 127.
%
\bibitem{gno} E. Belloti, talk at XIX International 
Conference on Neutrino Physics and Astrophysics, Sudbury, 
Canada, June 2000.
%
\bibitem{BKS}
J.~N.~Bahcall, P.~I.~Krastev and A.~Y.~Smirnov,
Phys.\ Lett.\  {\bf B477} (2000) 401; Phys.\ Rev.\ {\bf D 62} (2000) 093004.
%
\bibitem{valle}
M.C.~Gonzalez-Garcia, M.~Maltoni, C.~Pe\~{n}a-Garay and J.W.~Valle, 
Phys.\ Rev.\ {\bf D 63} (2001) 033005.
%
\bibitem{petcov}
M.~Maris and S.~T.~Petcov, Phys.\ Rev.\  {\bf D62} (2000) 093006.
%
\bibitem{LSND} C.~Athanassopoulos~\etal, LSND Collaboration,
Phys.\ Rev.\ {\bf C54} (1996) 2685; Phys.\ Rev.\ Lett.\ {\bf 77} (1996) 3082; 
Phys.\ Rev.\ Lett.\ {\bf 81} (1998) 1774.
%
\bibitem{bayo1}
M.~Fukugita and T.~Yanagida, Phys.\ Lett.\  {\bf B174} (1986) 45. 
%
\bibitem{bayo2}
W.~Buchm{\"u}ller and M.~Pl{\"u}macher, Phys.\ Lett.\  {\bf B389} (1996) 73; 
M.~Pl{\"u}macher, \hph{9807557}.
%
\bibitem{yanagidagroup}
T.~Asaka, K.~Hamaguchi, M.~Kawasaki and T.~Yanagida, Phys.\ Rev.\ {\bf D61}
(2000) 083512.
\bibitem{uppsala}
H.~B.~Nielsen and Y.~Takanishi,
Contributed to The Scandinavian Neutrino Workshop, Uppsala, Sweden, 
8-10 February 2001; \hph{0101181}.
%
\bibitem{NTbaryo}
H.~B.~Nielsen and Y.~Takanishi, to be appeared to 
Phys.\ Lett.\ {\bf B}; \hph{0101307}.
%
\bibitem{NTfutur}
H.~B.~Nielsen and Y.~Takanishi, in preparation.

\end{thebibliography}
\end{document}